\documentclass[prx,twocolumn,floatfix,superscriptaddress,longbibliography]{revtex4-1}

\usepackage{mathtools}

\usepackage{amssymb,amsmath,amstext,amsthm,mathtools}
\usepackage{graphicx}
\usepackage{epstopdf}
\usepackage{color}
\usepackage{bm}
\usepackage{appendix}
\usepackage[T1]{fontenc}
\usepackage{bbold}
\usepackage{bbm}
\usepackage{latexsym}
\usepackage[colorlinks=true,citecolor=blue,linkcolor=magenta]{hyperref}
\usepackage{float}
\usepackage{verbatim}
\usepackage{lipsum}
\usepackage{braket}
\usepackage{amsthm}

\newcommand{\HC}{\mathcal{H}}

\newcommand{\LC}{\mathcal{L}}

\renewcommand{\vec}[1]{\boldsymbol{#1}}  

\newcommand{\thv}{\vec{\theta}}

{}
{}

\theoremstyle{definition}

\begin{document}

\title{Challenges and Opportunities in Quantum Machine Learning}

\author{M. Cerezo}
\affiliation{Information Sciences, Los Alamos National Laboratory, Los Alamos, NM 87545, USA}
\affiliation{Center for Nonlinear Studies, Los Alamos National Laboratory, Los Alamos, New Mexico 87545, USA}
\affiliation{Quantum Science Center, Oak Ridge, TN 37931, USA}

\author{Guillaume Verdon}
\affiliation{X, Mountain View, CA, USA}
\affiliation{Institute for Quantum Computing, University of Waterloo, ON, Canada}
\affiliation{Department of Applied Mathematics, University of Waterloo, ON, Canada}

\author{Hsin-Yuan Huang}
\affiliation{Institute for Quantum Information and Matter, California Institute of Technology, USA}
\affiliation{Department of Computing and Mathematical Sciences, California Institute of Technology, USA}

\author{Lukasz Cincio}
\affiliation{Theoretical Division, Los Alamos National Laboratory, Los Alamos, New Mexico 87545, USA}
\affiliation{Quantum Science Center, Oak Ridge, TN 37931, USA}

\author{Patrick J. Coles}
\affiliation{Normal Computing Corporation, New York, New York, USA}
\affiliation{Theoretical Division, Los Alamos National Laboratory, Los Alamos, New Mexico 87545, USA}
\affiliation{Quantum Science Center, Oak Ridge, TN 37931, USA}

\begin{abstract}
At the intersection of machine learning and quantum computing, Quantum Machine Learning (QML) has the potential of accelerating data analysis, especially for quantum data, with applications for quantum materials, biochemistry, and high-energy physics. Nevertheless, challenges remain regarding the trainability of QML models. Here we review current methods and applications for QML. We highlight differences between quantum and classical machine learning, with a focus on quantum neural networks and quantum deep learning. Finally, we discuss opportunities for quantum advantage with QML.
\end{abstract}
\maketitle

\section{Introduction}\label{sc:intro}


 The recognition that the world is quantum mechanical has allowed researchers to embed well-established, but classical, theories into the framework of quantum Hilbert spaces. Shannon's information theory, which is the basis of communication technology, has been generalized to quantum Shannon theory (or quantum information theory), opening up the possibility that quantum effects could make information transmission more efficient~\cite{nielsen2000quantum}. The field of biology has been extended to quantum biology to allow for a deeper understanding of biological processes like photosynthesis, smell, and enzyme catalysis~\cite{brookes2017quantum}. Turing's theory of universal computation has been extended to universal quantum computation~\cite{deutsch1985quantum}, potentially leading to exponentially faster simulations of physical systems. 


One of the most successful technologies of this century is machine learning (ML), which aims to classify, cluster, and recognize patterns for large data sets. Learning theory has been simultaneously developed alongside of ML technology in order to understand and improve upon its success. Concepts like support vector machines, neural networks, and generative adversarial networks have impacted science and technology in profound ways. ML is now ingrained into society to such a degree that any fundamental improvement to ML  leads to tremendous economic benefit.


Like other classical theories, ML and learning theory can in fact be embedded into the quantum mechanical formalism. Formally speaking, this embedding leads to the field known as Quantum Machine Learning (QML)~\cite{wiebe2014quantumdeep,schuld2015introduction,biamonte2017quantum}, which aims to understand the ultimate limits of data analysis allowed by the laws of physics. Practically speaking, the advent of quantum computers, with the hope of achieving a so-called quantum advantage (as defined below) for data analysis, is what has made QML so exciting. Quantum computing exploits entanglement, superposition, and interference to perform certain tasks with significant speedups over classical computing, sometimes even exponentially faster. Indeed while such speedup has already been observed for a contrived problem~\cite{arute2019quantum}, reaching it for data science is still uncertain even at the theoretical level, but this is one of the main goals for QML.

\begin{figure}[t]
\centering
\includegraphics[width=.8\columnwidth]{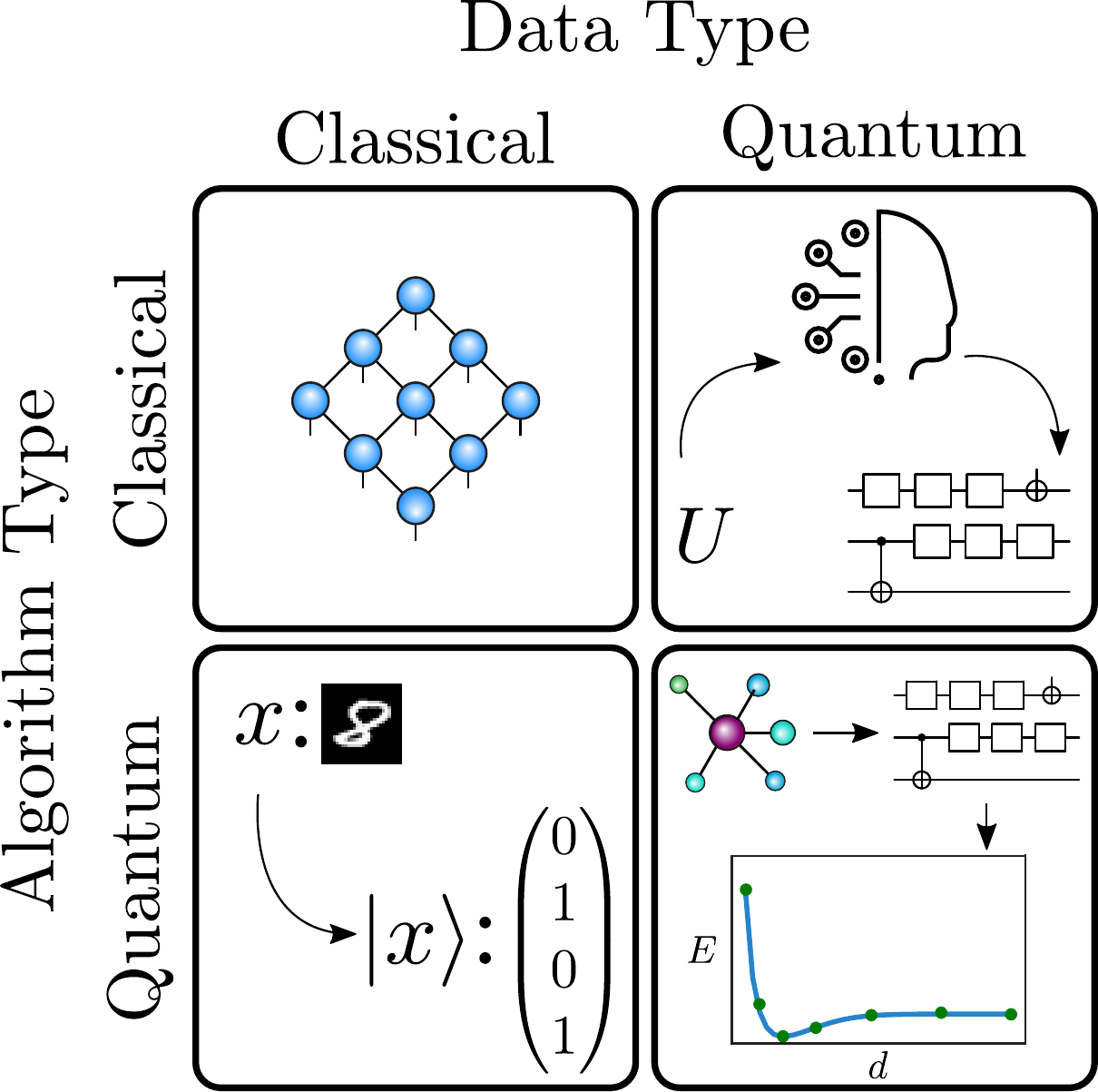}
\caption{\textbf{Quantum Machine Learning (QML) tasks}. Quantum machine learning is usually considered for four main tasks. These include tasks where the data is either classical or quantum, and where the algorithm is either classical or quantum. Top left: tensor networks are quantum-inspired classical methods that can analyze classical data. Top right: unitary time-evolution data $U$ from a quantum system can be classically compiled into a quantum circuit. Bottom left: handwritten digits can be mapped to quantum states for classification on a quantum computer. Bottom right: molecular ground state data can be classified directly on a quantum computer. The figure shows ground state energy $E$ dependence on the distance $d$ between the atoms.}
\label{fig:1}
\end{figure}

In practice, QML is a broad term that encompasses all of the tasks shown in Fig.~\ref{fig:1}. For example, one can apply machine learning to quantum applications like discovering quantum algorithms~\cite{cincio2018learning} or optimizing quantum experiments~\cite{tranter2018multiparameter,kaubruegger2021quantum}, or one can use a quantum neural network to process either classical or quantum information~\cite{cong2019quantum}. Even classical tasks can be viewed as QML when they are quantum inspired~\cite{tang2019quantum}. We note that the focus of this article will be on quantum neural networks, quantum deep learning, and quantum kernels, even though the field of QML is quite broad and goes beyond these topics.

\begin{figure}[t]
\centering
\includegraphics[width=1\columnwidth]{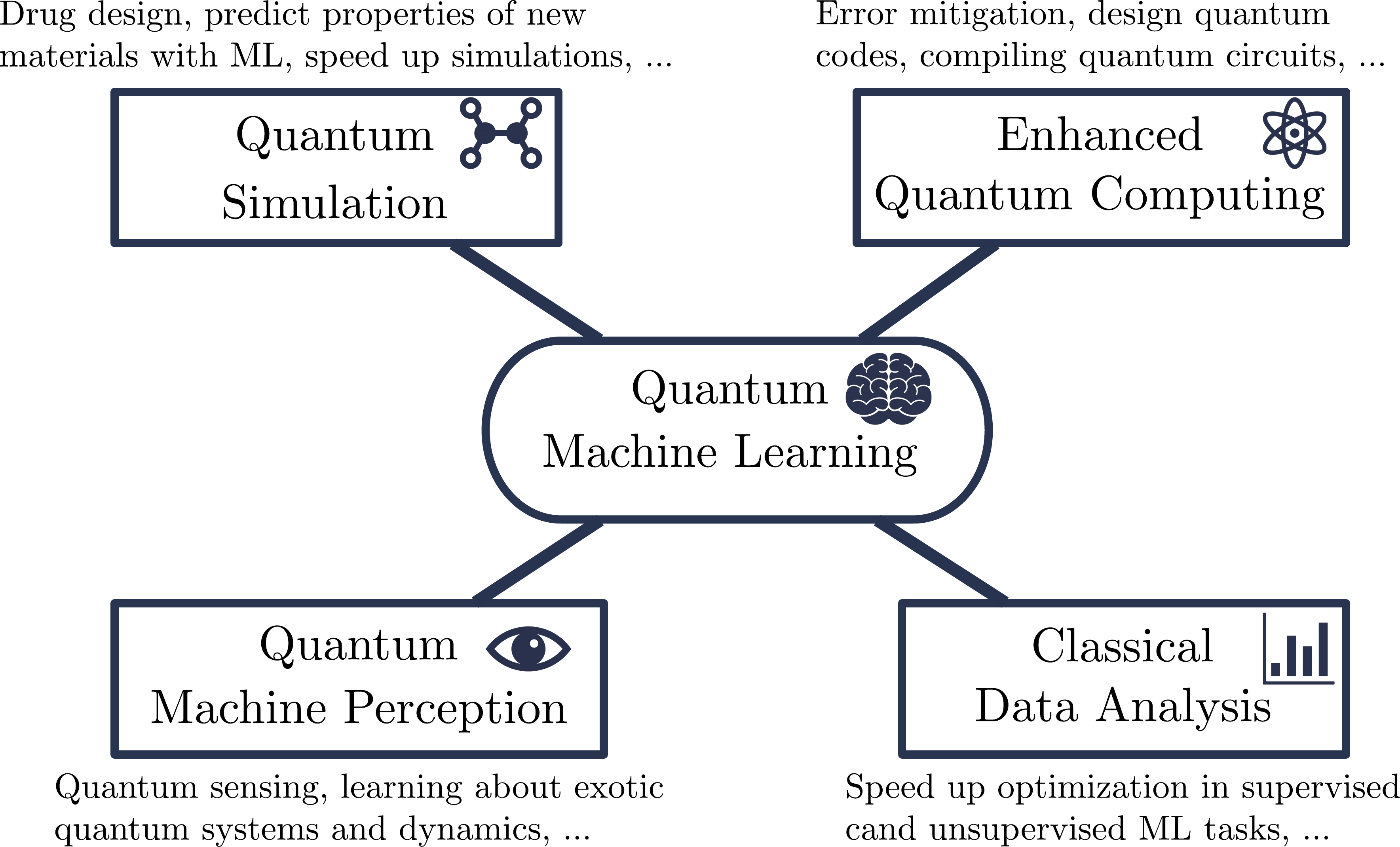}
\caption{\textbf{Key Applications for QML}. QML has been envisioned to bring a computational advantage in many applications. QML can enhance quantum simulation for chemistry (e.g., molecular ground states~\cite{peruzzo2014variational}, equilibrium states~\cite{verdon2019quantum}, and time evolution~\cite{cirstoiu2020variational}) and materials science (e.g., quantum phase recognition~\cite{cong2019quantum} and generative design with a target property in mind~\cite{sanchez2018inverse}). QML can enhance quantum computing by learning quantum error correction codes~\cite{cong2019quantum,johnson2017qvector} and syndrome decoders, performing quantum control, learning to mitigate errors, and compiling and optimizing quantum circuits. QML can enhance sensing and metrology~\cite{verdon2020qmp,meyer2020variational,beckey2020variational,broughton2020tensorflow,wang2017experimental} and extract hidden parameters from quantum systems. Finally, QML may speed up classical data analysis, including clustering and classification.}
\label{fig:applications}
\end{figure}

After the invention of the laser, it was called a solution in search of a problem. To some degree, the situation with QML is similar. The complete list of applications of QML is not fully known. Nevertheless, it is possible to speculate that all the areas shown in Fig.~\ref{fig:applications} will be impacted by QML. For example, QML will likely benefit chemistry, materials science, sensing and metrology, classical data analysis, quantum error correction, and quantum algorithm design. Some of these applications produce data that is inherently quantum mechanical, and hence it is natural to apply QML (rather than classical ML) to them.

While there are similarities between classical and quantum ML, there are also some differences. Because QML employs quantum computers, noise from these computers can be a major issue. This includes hardware noise like decoherence as well as statistical noise (i.e., shot noise) that arises from measurements on quantum states. Both of these noise sources can complicate the QML training process. Moreover, non-linear operations (e.g., neural activation functions) that are natural in classical ML require more careful design of QML models due to the linearity of quantum transformations.


For the field of QML, the immediate goal for the near-future is demonstrating quantum advantage, i.e., outperforming classical methods, in a data science application. Achieving this goal will require keeping an open mind about which applications will benefit most from QML (e.g., it may be an application that is inherently quantum mechanical). Understanding how QML methods scale to large problem sizes will also be required, including analysis of trainability (gradient scaling) and prediction error. The availability of high quality quantum hardware~\cite{huang2021power,banchi2021generalization} will also be crucial.


Finally, we note that QML provides a new way of thinking about established fields, like quantum information theory, quantum error correction, and quantum foundations. Viewing such applications from a data science perspective will likely lead to new breakthroughs.


\section{Framework}

\subsection{Data}

\begin{figure}[t]
\centering
\includegraphics[width=1\columnwidth]{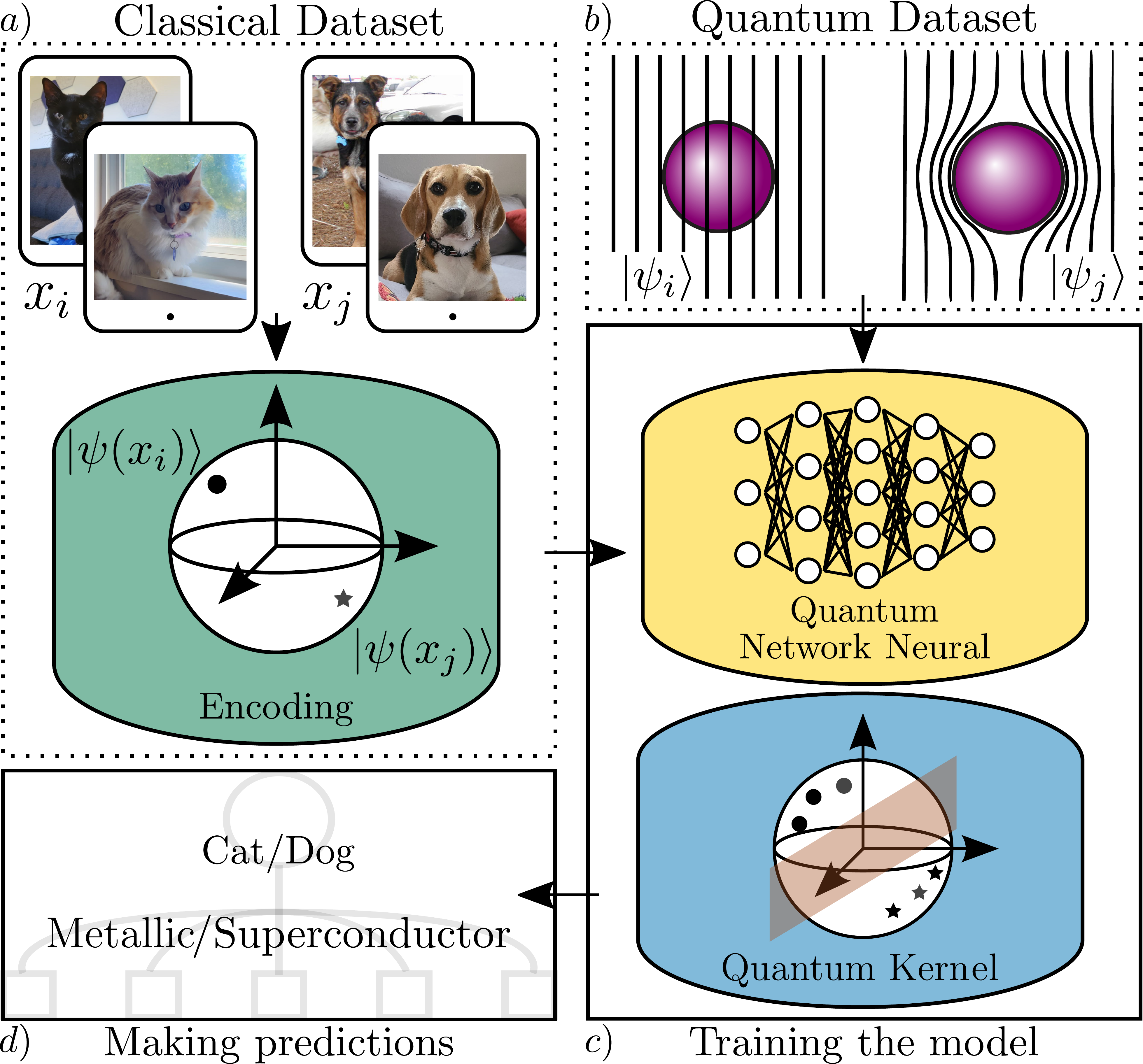}
\caption{ \textbf{Classification with QML.} a) The classical data $x$, i.e., images of cats and images of dogs, is  encoded into a Hilbert space via some map $x\rightarrow\ket{\psi(x)}$. Ideally, data from different classes (here represented by dots and stars)  is mapped to different regions of the Hilbert space . b) Quantum data $\ket{\psi}$ can be directly analyzed on a quantum device. Here the dataset is composed of states representing metallic  or superconducting systems.  c) The dataset is used to train a QML model. Two common paradigms in QML are quantum neural networks and quantum kernels, both of which allow for classification of either classical or quantum data. In Kernel methods one fits a decision hyperplane that separates the classes.  d) Once the model is trained, it can be used to make predictions.  }
\label{fig:QMLsetting}
\end{figure}

As shown in Fig.~\ref{fig:QMLsetting},  QML can be used to learn from either classical or quantum data, and thus we begin by  contrasting these two types of data. Classical data is ultimately encoded in bits, each of which can be in a  $0$ or  $1$ state. This includes images, texts, graphs, medical records, stock prices, properties of molecules, outcomes from biological experiments, and collision traces from high energy physics experiments. Quantum data is encoded in quantum bits, called qubits, or higher-dimensional analogs. A qubit can be represented by the states $\ket{0}$, $\ket{1}$, or any normalized complex linear \textit{superposition} of these two. Here, the states contain information obtained from some physical process such as quantum sensing~\cite{degen2017quantum}, quantum metrology~\cite{giovannetti2011advances}, quantum networks~\cite{chiribella2009theoretical}, quantum control~\cite{dalessandro2010introduction}, or even quantum analog-digital transduction~\cite{verdon2020qadi}. Moreover, quantum data can also be the solution to problems obtained on a quantum computer, e.g., the preparation of various Hamiltonians' ground states. 

In principle, all classical data can be efficiently encoded in systems of qubits: a classical bitstring of length $n$ can be easily encoded onto $n$ qubits. However, the same cannot be said for the converse, since one cannot efficiently encode quantum data in bit systems, i.e., the state of a general $n$ qubit system requires $(2^n-1)$ complex numbers to be specified. Hence, systems of qubits (and more generally the quantum Hilbert space) constitute the ultimate data  representation medium, as they can encode not only classical information but also quantum information obtained from physical processes.

In a QML setting, the term quantum data refers to data that is naturally already embedded in a Hilbert space $\HC$. When the data is quantum, it is already in the form of a set of quantum states $\{\ket{\psi_j}\}$ or a set of unitaries $\{U_j\}$ that could prepare these states on a quantum device (via the relation $\ket{\psi_j} = U_j\ket{\vec{0}}$). On the other hand, when the data $x$ is classical, it first needs to be encoded in a quantum system through some embedding mapping $x_j\rightarrow \ket{\psi(x_j)}$, with $\ket{\psi(x_j)}$ in $\HC$. In this case, the hope is that the QML model can solve the learning task by accessing the exponentially large dimension of the Hilbert space~\cite{rebentrost2014quantum,schuld2019quantum,lloyd2020quantum,schuld2021effect}.

One of the most important and reasonable conjectures to make is that the availability of quantum data will significantly increase in the near future. The mere fact that people will use the quantum computers that are available will logically lead to more quantum problems being solved and quantum simulations being performed. These computations will produce quantum data sets, and hence it is reasonable to expect the rapid rise of quantum data.  Note that, in the near term, this quantum data will be stored on classical devices in the form of efficient descriptions of quantum circuits that prepare the datasets.

Finally, as our level of control over quantum technologies progress, coherent transduction of quantum information from the physical world to digital quantum computing platforms may be achieved \cite{verdon2020qadi}. This would quantum mechanically mimic the main information acquisition mechanism for classical data from the physical world, that being analog-digital conversion. Moreover, we can expect that the eventual advent of practical quantum error correction~\cite{roffe2019quantum} and quantum memories~\cite{shor1995scheme} will allow us to store quantum data on quantum computers themselves.

\subsection{Models}


Analyzing and learning from data requires a parameterized model, and many different models have been proposed for QML applications. Classical models like neural networks and tensor networks (as shown in Fig.~\ref{fig:1}) are often useful for analyzing data coming from quantum experiments. However, due to their novelty, we will focus our discussion on quantum models using quantum algorithms, where one applies the learning methodology directly at the quantum level. 

Similar to classical ML, there exists several different QML paradigms: supervised learning (task-based)~\cite{havlivcek2019supervised,liu2021rigorous,schuld2021quantum}, unsupervised learning (data-based)~\cite{otterbach2017unsupervised,kerenidis2019q} and reinforced learning (reward-based)~\cite{saggio2021experimental,skolik2021quantum}. While each of these fields is exciting and thriving on its own, supervised learning has recently received considerable attention for its potential to achieve quantum advantage~\cite{huang2021quantumadvantage,havlivcek2019supervised}, resilience to noise~\cite{larose2020robust}, and good generalization properties~\cite{caro2021generalization,caro2022outofdistribution,caro2021encodingdependent}, which makes it a strong candidate for near-term applications. In what follows we discuss two popular QML models: quantum neural networks (QNNs) and quantum kernels, shown in Fig.~\ref{fig:QMLsetting}, with a particular emphasis on QNNs as these are the primary ingredient of several supervised, unsupervised, and reinforced learning schemes.

\subsubsection{Quantum neural networks}

The most basic and key ingredient in QML models are Parameterized Quantum Circuits (PQCs). These involve a sequence of unitary gates acting on the quantum data states $\ket{\psi_j}$, some of which have free parameters $\thv$ that will be trained to solve the problem at hand~\cite{cerezo2020variationalreview}.   PQCs are conceptually analogous to neural networks, and indeed this analogy can be made precise, i.e., classical neural networks can be formally embedded into PQCs~\cite{wan2017quantum}.

This has led researchers to refer to certain kinds of PQCs as Quantum Neural Networks (QNNs). In practice, the term QNN is used whenever a PQC is employed for a data science application, and hence we will use the term QNN in what follows. QNNs are employed in all three QML paradigms mentioned above. For instance, in a supervised classification task, the  goal of the QNN is to map the states in different classes to distinguishable regions of the Hilbert space~\cite{havlivcek2019supervised}. Moreover, in the unsupervised learning scenario of~\cite{otterbach2017unsupervised} a clustering task is mapped onto a MAXCUT problem and solved by training a QNN to maximize distance between classes. Finally, in the reinforced learning task of~\cite{skolik2021quantum}, a QNN can be used as the Q-function approximator, which can be used to determine the best action for a learning agent given its current state.

\begin{figure}
\centering
\includegraphics[width=1\columnwidth]{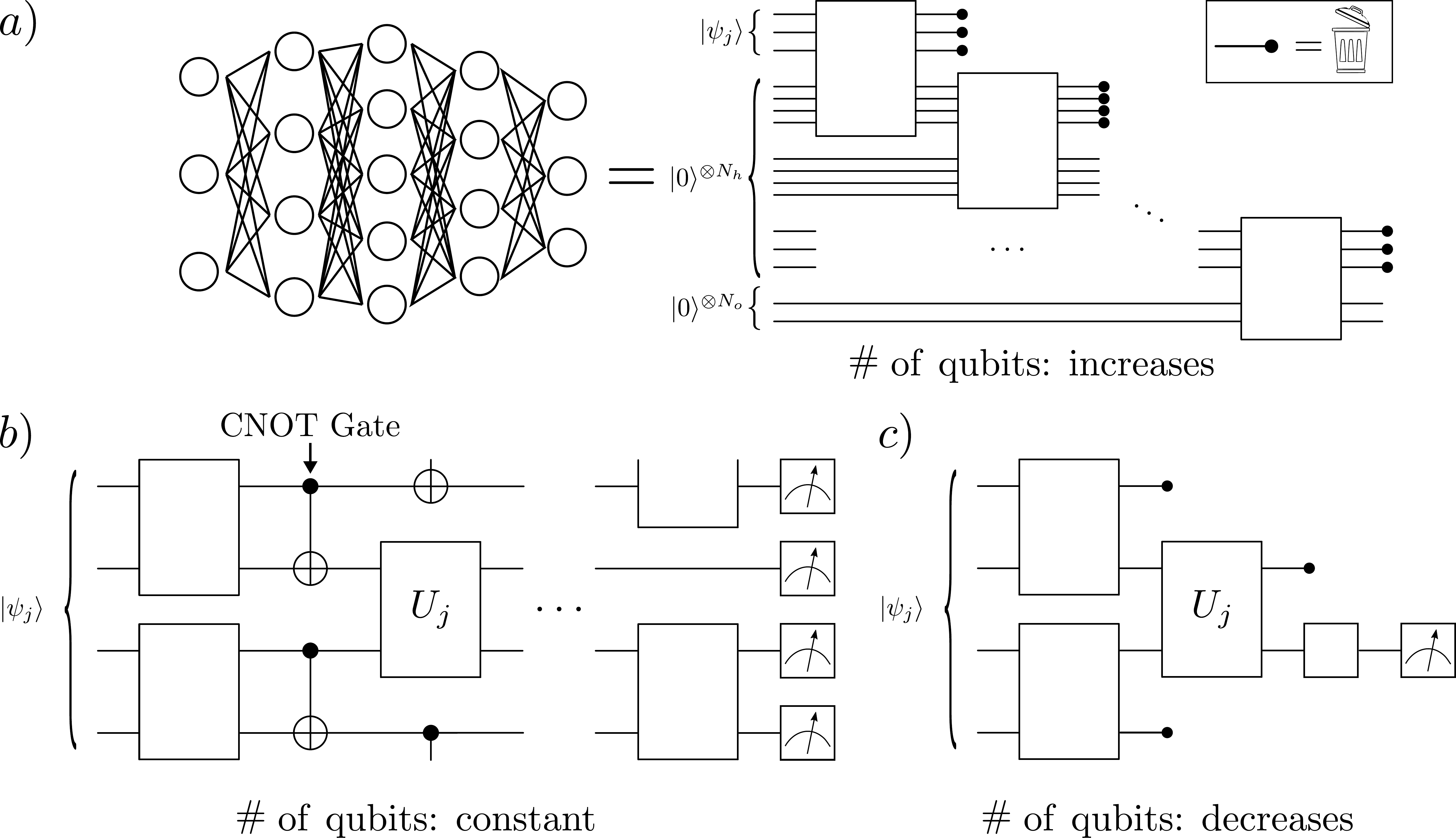}
\caption{ \textbf{Examples of QNN architectures.} a) A classical feed-forward neural network has input, hidden, and output layers. This can be generalized to the quantum setting with a dissipative QNN, where some qubits are discarded and replaced by new qubits during the algorithm . Here we shown a quantum circuit representation for the dissipative QNN. In a circuit diagram each horizontal line represents a qubit, and the logical operations, or quantum gates, are represented by boxes connecting the qubit lines. Circuits are read from left to right. For instance, here the circuit is initialized in a product state $\ket{\psi_j} \otimes \ket{0}^{\otimes (N_h + N_o)}$, where $\ket{\psi_j}$ encodes the input data and $N_h$ ($N_o$) is the number of qubits in the hidden (output) layer. As one performs logical operations, the information forward propagates through the circuit. b) Another possible QNN strategy is to keep the qubits fixed, without discarding or replacing them. The circuit represents consecutive application of two-qubit gates $U_j$ and controlled-NOT (denoted by CNOT) gates. c) Quantum convolutional neural networks (QCNNs) measure and discard qubits during the algorithm. The QCNN circuit considered here is built with two-qubit quantum gates $U_j$ and is initialized in $\ket{\psi_j}$. }

\label{fig:QNNexamples}
\end{figure}

Figure~\ref{fig:QNNexamples} gives examples of three distinct QNN architectures where at each layer the number of qubits in the model is increased, preserved, or decreased. In Fig.~\ref{fig:QNNexamples}(a) we show a  dissipative QNN~\cite{beer2020training} which generalizes the classical feed-forward network. Here, each node corresponds to a qubit, while lines connecting qubits are unitary operations.  The term dissipative arises from the fact that qubits in a layer are discarded after the information forward-propagates to the (new) qubits in the next layer. Figure~\ref{fig:QNNexamples}(b) shows a standard  QNN where quantum data states are sent through a quantum circuit, at the end of which some or all of the qubits are measured. Here, no qubits are discarded or added as one goes deeper into the QNN. Finally, Fig.~\ref{fig:QNNexamples}(c) depicts a convolutional QNN~\cite{cong2019quantum}, where at each layer qubits are measured to reduce the dimension of the data while preserving its relevant features. Many other QNNs have been proposed~\cite{schuld2014quest,dallaire2018quantum,farhi2018classification,killoran2019continuous,bausch2020recurrent}, and constructing QNN architectures is currently an active area of research.

To further accommodate for the limitation of near-term quantum computers, one can also employ a hybrid approach with models that have both classical and quantum neural networks~\cite{broughton2020tensorflow}. Here, QNNs act coherently on quantum states while deep classical neural networks  alleviate the need for higher-complexity quantum processing. Such hybridization  distributes the representational capacity and computational complexity across both quantum and classical computers. Moreover, since quantum states generally have a mixture of classical correlations and quantum correlations, hybrid quantum-classical models allow for the use of quantum computers as an additive resource to increase the ability for classical models to represent quantum-correlated distributions.   Applications of hybrid models include generating~\cite{verdon2019quantum} or learning and distilling information~\cite{broughton2020tensorflow} from multipartite-entangled distributions.

\subsubsection{Quantum kernels}

As an alternative to QNNs, researchers have proposed quantum versions of kernel methods~\cite{havlivcek2019supervised,schuld2021quantum}.
A kernel method maps each input to a vector in a high-dimensional vector space, known as the reproducing kernel Hilbert space.
Then, a kernel method learns a linear function in the reproducing kernel Hilbert space.
The dimension of the reproducing kernel Hilbert space could be infinite, which enables the kernel method to be very powerful in terms of the expressiveness.
To learn a linear function in a potentially infinite-dimensional space, the kernel trick \cite{cortes1995support} is employed, which only requires efficient computation of the inner product between these high-dimensional vectors.
The inner product is also known as the kernel \cite{cortes1995support}.
Quantum kernel methods consider the computation of kernel functions using quantum computers. There are many possible implementations. For example, \cite{havlivcek2019supervised, schuld2021quantum} considered a reproducing kernel Hilbert space equal to the quantum state space, which is finite dimensional. Another approach~\cite{huang2021power} is to 
study an infinite-dimensional reproducing kernel Hilbert space that is equivalent to transforming classical vector using a quantum computer. It  then maps the transformed classical vectors to infinite-dimensional vectors.

\subsubsection{Inductive bias}

For both QNNs and quantum kernels, an important design criterion is their inductive bias. This bias refers to the fact that any model represents only a subset of functions and is naturally biased towards certain types of functions (i.e, functions relating the input features to the output prediction). One aspect of achieving quantum advantage with QML is to aim for QML models that have an inductive bias that is inefficient to simulate with a classical model. Indeed, it was recently shown~\cite{kubler2021inductive} that quantum kernels with this property can be constructed, albeit with some subtleties regarding their trainability. 


Generally speaking, inductive bias encompasses any assumptions made in the design of the model or the optimization method which bias the search of the potential models to a subset in the set of all possible models. In the language of Bayesian probabilistic theory, we usually call these assumptions our prior. Having a certain parameterization of potential models, like QNNs, or choosing a particular embedding for quantum kernel methods \cite{havlivcek2019supervised,huang2021power,banchi2021generalization} is itself a restriction of the search space, and hence a prior. Adding a regularization term to the optimizer or modulating the learning rate to keep searches geometrically local also adds inherently a prior and focuses the search, and thus provides inductive bias.

Ultimately, inductive biases from the design of the ML model, combined with a choice of training process, are what make or break an ML model. The main advantage of QML will then be to have the ability to sample from and learn models that are (at least partially) natively quantum mechanical. As such, they have inductive biases that classical models do not have. This discussion assumes that the dataset to be represented is quantum mechanical in nature, and is one of the reasons why researchers typically believe that QML has greater promise from quantum rather than classical data.

\subsection{Training and Generalization}

The ultimate goal of ML (classical or quantum) is to train a model to solve a given task. Thus, understanding the training process of QML models is fundamental for their success.

Consider the training process, whereby one aims to find the set of parameters $\thv$ that lead to the best performance. The latter can be accomplished, for instance, by minimizing a loss function $\LC(\thv)$ encoding the task at hand. Some methods for training QML models are leveraged from classical ML, like stochastic gradient descent. However, shot noise, hardware noise, and unique landscape features often make off-the-shelf classical optimization methods perform poorly for QML training~\footnote{This is due to the fact that extracting information from a quantum state requires computing the expectation values of some observable, which in practice need to be estimated via measurements on a noisy quantum computer. Hence, given a finite number of shots (measurement repetitions), these can only be resolved up to some additive errors. Moreover, such expectation values will be subject to corruption due to hardware noise.}. This realization led to development of quantum-aware optimizers, which account for the quantum idiosyncrasies of the QML training process. For example, shot-frugal optimizers~\cite{kubler2020adaptive, arrasmith2020operator, gu2021adaptive, sweke2020stochastic} can employ stochastic gradient descent while adapting the number of shots (or measurements) needed at each iteration, so as not to waste too many shots during the optimization. Quantum natural gradient~\cite{stokes2020quantum,koczor2019quantum} adjusts the step size according to the local geometry of the landscape (based on the quantum Fisher information metric). These and other quantum-aware optimizers often outperform standard classical optimization methods in QML training tasks.

For the case of supervised learning, one is not only interested in learning from a training data set but also in making accurate predictions on (generalize to) previously unseen data. This translates into achieving small training and prediction errors, with the second usually hinging on the first.  Thus, let us now consider prediction error, also known as generalization error, which has been studied only very recently for QML~\cite{sharma2020reformulation,abbas2020power,huang2021power,banchi2021generalization,caro2021encodingdependent,caro2021generalization}. Formally speaking, this error measures the extent to which a trained QML model performs well on unseen data. Prediction error depends on both the training error as well as the complexity of the trained model. If the training error is large, the prediction error is also typically large. If the training error is small but the complexity of the trained model is large, then the prediction error is likely still large. The prediction error is small only if training error is itself small and the complexity of the trained model is moderate (i.e., sufficiently smaller than training data size)~\cite{caro2021generalization,banchi2021generalization}.
The notion of complexity depends on the QML model. We have a good understanding of the complexity of quantum kernel methods~\cite{huang2021power,banchi2021generalization}, while more research is needed on QNN complexity.
Recent theoretical analysis of QNNs shows that their prediction performance is closely linked to the number of independent parameters in the QNN, with good generalization obtained when the amount of training data is roughly equal to the number of parameters~\cite{caro2021generalization}. This gives the exciting prospect of using only a small amount of training data to obtain good generalization. 

\section{Challenges in QML}


\begin{figure}
\centering
\includegraphics[width=1\columnwidth]{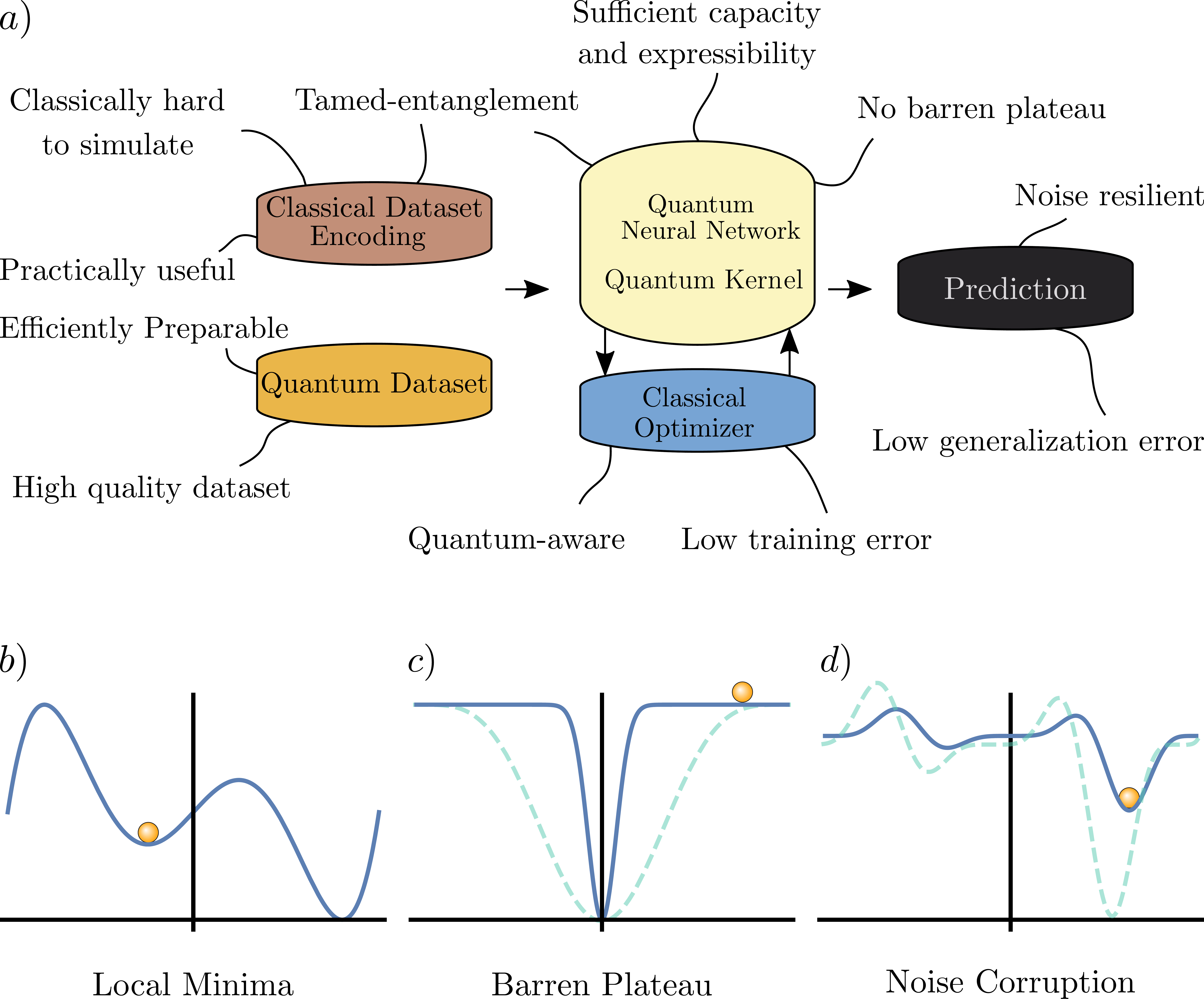}
\caption{\textbf{Challenges for QML.} a) There are several ingredients and  priors needed to build a QML model:  a dataset (and an encoding scheme for classical data), the choice of parameterized model, loss function, and classical optimizer.  In this diagram, we show some of the challenges of the different components of the model. b-d) The success of the QML model hinges on an accurate and efficient training of the parameters. However, there are certain phenomena that can hinder the QML trainability. These include the abundance of low-quality local minima solutions shown in b), as well as the barren plateau phenomenon in c).  When a QML architecture exhibits a barren plateau, the landscape becomes exponentially flat (on average) as the number of qubits increases (seen as a transition from dashed to solid line). The presence of hardware noise has been shown to erase the features in the landscape as well as potentially shift the position of the minima. Here, the dashed (solid) line corresponds to the noiseless (noisy) landscape shown in d).}
\label{fig:5}
\end{figure}

Heuristic fields can face periods of stagnation (or ``winters'') due to unforeseen technical challenges. Indeed in classical ML, there was a gap between introducing a single perceptron~\cite{rosenblatt1957perceptron} and the multi-layer perceptron~\cite{haykin1994neural} (i.e., neural network), and there was also a gap between attempts to train multiple layers and the introduction of the backpropagation method~\cite{rumelhart1986learning}.

Naturally we would like to avoid these stagnations or winters for QML. The obvious strategy is to try to determine all of the challenges as quickly possible, and focus research effort on addressing them. Fortunately, QML researchers have taken this strategy. Figure~\ref{fig:5} showcases some  of the different elements of QML models, as well as the challenges associated with them. In this section we detail various QML challenges and how one could potentially avoid them. 

\subsection{Embedding schemes and quantum datasets}

The access to high-quality, standardized datasets has played a key role in advancing classical ML. Hence, one could conjecture that such datasets will be crucial for QML as well. 

Currently, most QML architectures are benchmarked using classical datasets (such as MNIST, Dogs vs Cats, and Iris). While using classical datasets is natural due to their accessibility, it is still unclear how to best encode classical information onto quantum states. Several embedding schemes have been proposed~\cite{havlivcek2019supervised,lloyd2020quantum,hubregtsen2021training}, and there are some desirable properties they must possess. One such property is that the inner product between  output states of the embedding is classically hard to simulate (otherwise the quantum kernel would be classically simulable). In addition, the embedding should be practically useful, i.e., in a classification task, the states should be in  distinguishable regions of the Hilbert space. Unfortunately, embeddings that satisfy one of these properties do not necessarily satisfy the other~\cite{thanasilp2021subtleties}. Thus, developing  encoding schemes  is an active area of research, especially those that are equipped with an  inductive bias containing information about the dataset~\cite{kubler2021inductive}. 

Furthermore, some recent results suggest that achieving a quantum advantage with  classical data might not be straightforward~\cite{kubler2021inductive}. On the other hand, QML models with quantum data have a more promising route towards a quantum advantage~\cite{huang2021quantum,cotler2021revisiting, chen2021hierarchy, chen2021exponential}. Despite this fact, there is still a dearth of truly quantum datasets for QML, which just a few recently proposed~\cite{perrier2021qdataset,schatzki2021entangled}. Hence, the field needs standardized quantum datasets with easily preparable quantum states, as these can be used to benchmark QML models on true quantum data.

\subsection{Quantum landscapes}

Training the parameters of the QML model  corresponds in a wide array of cases to minimizing a loss function and navigating through a (usually non-convex) loss function landscape in search for its global minimum~\footnote{Technically speaking, the loss function defines a map from the model's parameter space to the real values. The loss function value can quantify, for instance, the model's error in solving a given so that our goal is to find the set of parameters that minimizes such error.}. Quantum landscape theory~\cite{arrasmith2021equivalence} aims to understand QML landscape properties and how to engineer them. Local minima and barren plateaus have received significant attention in quantum landscape theory.

\subsubsection{Local minima in quantum landscapes}

As schematically shown in Fig.~\ref{fig:5}(b), similar to classical ML,   the quantum loss landscape can have many local minima. Ultimately, this can lead to the overall non-convex optimization being NP-hard~\cite{bittel2021training}, which is again similar to the classical case. There have been some methods proposed to address local minima. For example, variable structure QNNs~\cite{bilkis2021semi,larose2019variational}, which grow and contract throughout the optimization, adaptively  change the model's prior and allow some local minima to be turned into saddle points. Moreover, evidence of the overparametrization phenomenon has been seen for QML~\cite{kiani2020learning,larocca2021theory}. Here, the optimization undergoes a computational phase transition, due to spurious local minima disappearing, whenever the number of parameters exceeds a critical value.

\subsubsection{Overview of barren plateaus}

Local minima are not the only issue facing QML, as it has been shown that quantum landscapes can exhibit a  fascinating property known as a  \textit{barren plateau}~\cite{mcclean2018barren,cerezo2021cost,cerezo2020impact,arrasmith2020effect,holmes2021connecting,pesah2020absence,volkoff2021large,sharma2020trainability,holmes2020barren,marrero2020entanglement,uvarov2020barren,patti2020entanglement,abbas2020power,wang2020noise}. As depicted in Fig.~\ref{fig:5}(c), in a barren plateau the loss landscape becomes, on average, exponentially flat with the problem size. When this occurs, the valley containing the global mimimum also shrinks exponentially with problem size, leading to a so-called \textit{narrow gorge}~\cite{arrasmith2021equivalence}. As a consequence, one requires exponential resources (e.g., numbers of shots) to navigate through the landscape. The latter impacts the complexity of one's QML algorithm and can even destroy quantum speedup, since quantum algorithms typically aim to avoid the exponential complexity normally associated with classical algorithms.

\subsubsection{Barren plateaus from ignorance\\ or insufficient inductive bias}

The barren plateau phenomenon was  first studied in deep hardware-efficient QNNs~\cite{mcclean2018barren}, where they arise due to the high \textit{expressivity} of the model~\cite{holmes2021connecting}. By making no  assumptions about the underlying data, deep hardware-efficient architectures aims to solve a problem by being able to prepare a wide range of unitary evolutions. In other words, the prior over hypothesis space is relatively uninformed. Barren plateaus in this unsharp prior are caused by ignorance or the lack of sufficient inductive bias, and therefore a means to avoid them is to input knowledge into the construction of the QNN - making the design of QNNs with good inductive biases for the problem at hand a key solution.  

Fortunately various strategies have been developed to address these barren plateaus, such as clever initialization~\cite{verdon2019learning}, pre-training, and parameter correlation~\cite{volkoff2021large,pesah2020absence}. These are all examples of adding a sharper prior to one's search over the over-expressive parameterizations of hardware efficient QNNs. Below we further discuss how QNN architectures can be designed to further introduce inductive bias.

\subsubsection{Barren plateaus from global observables}

Other mechanisms have been linked to barren plateaus. Simply defining a loss function based on a global observable (i.e., observables measuring all qubits) leads to barren plateaus even for shallow circuits with sharp priors~\cite{cerezo2021cost}, while local observables (those comparing quantum states at the single-qubit level) avoid this issue~\cite{cerezo2021cost, uvarov2020barren}. The latter is not due to bad inductive biases but rather to the fact that comparing objects in exponentially large Hilbert spaces requires an exponential precision, as their overlap is usually exponentially small. 

\subsubsection{Barren plateaus from entanglement}

While entanglement is one of the most important quantum resources for information processing tasks in quantum computers, it can also be detrimental for QML models. QNNs (or embedding schemes) that generate too much entanglement  also  lead to barren plateaus~\cite{sharma2020trainability,marrero2020entanglement,patti2020entanglement}.  Here, the issue arises when one entangles the visible qubits of the QNN (those that one measures at the QNN's output) with a large number of qubits in the hidden layers. Due to entanglement, the information of the state is stored in non-local correlations across all qubits, and hence the reduced state of the visible qubits concentrates around the maximally mixed state. This type of barren plateau can be solved by taming the entanglement generated across the QNN.

\subsection{QNN architecture design}

One of the most active areas  is developing QNN architectures  that have sharp  priors. Since QNNs are a fundamental ingredient in supervised learning (deep learning, kernel methods), but also in unsupervised learning and reinforced learning, developing good QNN architectures is crucial for the field. 

For instance, it has been shown that QNNs with sharp priors can  avoid issues
such as barren plateaus altogether. One such example are Quantum Convolutional Neural Networks (QCNNs)~\cite{cong2019quantum}.  QCNNs possess an inductive bias from having a prior over the space of architectures that is much sharper than that of deep hardware-efficient architectures, as QCNNs are restricted to be hierarchically structured and translationally invariant. The significant reduction in the expressivity and parameter space dimension from this translational invariance assumption yields the greater trainability~\cite{pesah2020absence}.

The idea of embedding knowledge about the problem and dataset into our models (to achieve helpful inductive bias) will be key to improve the trainability of QML models. Recent proposals use Quantum Graph Neural Networks~\cite{verdon2019quantumgraph} for scenarios where quantum subsystems live on a graph, and potentially have further symmetries. For instance, the underlying graph-permutation symmetries of a quantum communication dataset were taken into account by a quantum graph convolutional network. Similarly, a quantum recurrent neural network has been used in scenarios where temporal recurrence of parameters occurs, e.g., as in the quantum dynamics of a stationary (time-dependent) quantum dynamical process.

To better understand how to go beyond the aforementioned inductive biases from temporal and/or translational invariance in grids and graphs, we can take inspiration from recent advances in the theory of classical deep learning. In classical ML, the study of the group theory  behind graph neural networks, namely the concepts of invariance and equivariance to various group actions on the input space, has led to a unifying theory of deep learning architectures based on group theory, called Geometric Deep Learning theory \cite{bronstein2021geometric}.

In order to have a prescription to create arbitrary architectures and inductive biases suitable for a given set of quantum physical data, a theory of quantum geometric deep learning could be key to design architectures with the right prior over the transformation space and inductive biases to ensure trainability and generalization. As the study of physics is often about the identification of inherent or emergent symmetries in particular systems, there is great potential for a future unifying theory of quantum geometric deep learning to provide consistent methods to create QML model architectures with inductive biases encoding knowledge of the basic symmetries and principles of the quantum physical system underlying given quantum datasets. This approach has been recently explored in~\cite{larocca2022group,skolik2022equivariant,meyer2022exploiting}. Moreover, the works of~\cite{larocca2021diagnosing,larocca2021theory} have also shown that the Lie algebra obtained from the generators of the QNN can be linked to properties of the QML landscape such as the presence of barren plateaus or the overparametrization phenomenon.

\subsection{Effect of quantum noise}

The presence of hardware noise during quantum computations is one of the defining characteristics of Noisy Intermediate-Scale Quantum (NISQ) computing. Despite this fact, most QML research neglects noise in the analytical calculations and numerical simulations while still promising that the methods are near-term compatible. Accounting for the effects of hardware noise should be a crucial aspect of QML analysis if one wishes to pursue a quantum advantage with currently available hardware. 

Noise corrupts the information as it forward propagates in a quantum circuit, meaning that deeper circuits with longer run-times will be particularly affected. As such, noise affects all aspects of the model that make use of quantum  computers. This includes the dataset preparation scheme as well as  circuits used to compute quantum kernels. Moreover, when using QNNs, noise can hinder their trainability as it leads to noise-induced barren plateaus~\cite{wang2020noise,wang2021can}. Here, the relevant features of the landscape get exponentially suppressed by noise as the  depth of the circuit increases (see Fig.~\ref{fig:5}(d)). Ultimately, the effects of noise translate into a deformation of the inductive bias of the model from its original one, and an effective reduction of the dimension of the quantum feature space. Despite the critical impact of quantum noise, its effects are still largely unexplored, particularly on its impact on the classical simulability of the QML model~\cite{deshpande2021tight,hakkaku2021quantifying}.

Addressing noise-induced issues will likely require either: (1) reduction in hardware error rates, (2) partial quantum error correction~\cite{bultrini2022battle}, or (3) employing QNNs that are relatively shallow (i.e., whose depth grows sublinearly in the problem size)~\cite{wang2020noise}, such as QCNNs. Error mitigation techniques~\cite{temme2017error,czarnik2020error,endo2021hybrid} can also improve performance of QML models in the presence of noise, although they may not solve noise-induced trainability issues~\cite{wang2021can}. A different approach to dealing with noise is to engineer  QML models with noise-resilient properties~\cite{sharma2019noise,larose2020robust,cincio2020machine} (such as the position of the minima not changing due to noise).

\section{Outlook}

\subsection{Potential for Quantum advantage}

The first quantum advantages in QML will likely come from hidden parameter extraction from quantum data. This can be for quantum sensing or quantum state classification/regression. Fundamentally, we know from the theory of optimal measurement that non-local quantum measurements can extract hidden parameters using less samples. Using QML, one can form and search over a parameterization of hypotheses for such measurements. 

This is particularly useful when such optimal measurements are not known \textit{a priori}, for example, identifying the measurement that extracts an order parameter or identifies a particular phase of matter. As the information about this classical parameter is embedded in the structure of quantum correlations between subsystems, it is natural that a trained QML model with good inductive biases can exhibit an advantage over local measurements and classical representations.

Another area of application where classical parameter extraction may yield an advantage is in quantum machine perception ~\cite{verdon2020qmp,meyer2020variational,beckey2020variational,broughton2020tensorflow,wang2017experimental,huang2021quantum}, i.e. quantum sensing, metrology, and beyond. Here, leveraging the variational search over multipartite-entangled states for input to exposure to a quantum signal along with the optimization for optimal control and/or over post-processing schemes can find optimal measurements for the estimation of hidden parameters in the incoming signal. In particular, the variational approach may be able to find the optimal entanglement, exposure, and measurement scheme which filters signal from the noise \cite{layden2018spatial}, akin to variationally learning the quantum error correcting code which filters signal from noise, instead applied to quantum metrology.

Beyond classical parameter extraction embedded in quantum data, there may be an advantage for the discovery of quantum error correcting codes (QECCs) \cite{johnson2017qvector}. QECC's fundamentally encode data (typically) non-locally into a subsystem or subspace of the Hilbert space. As deep learning is fundamentally about the discovery of submanifolds of data space, identifying and decoding subspaces/subsystems from a Hilbert space which correspond to a quantum error correction subspace/subsystem is a natural place where differentiable quantum computing may yield an advantage. This is a barely explored area, mainly due to the difficulty of gaining insights with small-scale numerical simulations. Fundamentally, it is akin to a quantum data version of classical parameter embedding/extraction advantage.

Finally, a quantum advantage for generative modelling may be achieved when one can generate ground states \cite{peruzzo2014variational}, equilibrium states \cite{mcardle2019variational,verdon2019quantum}, or quantum dynamics \cite{cirstoiu2020variational}, using generative models incorporating QNNs, in a way where the distribution cannot be sampled classically, and yields more accurate predictions or more extensively generalization compared to classical ML approaches. The nearest-term possibility for demonstrating such an advantage would likely be from variational optimization at the continuous time optimal control level on analogue quantum simulators.

\subsection{What will quantum advantage look like?}

When the data originates from quantum-mechanical processes, such as from experiments in chemistry, material science, biology, and physics, it is more likely to see exponential quantum advantage in ML. The quantum advantage could be in \emph{sample complexity} or \emph{time complexity}. An exponential advantage in sample complexity always implies an exponential advantage in time complexity, but the reverse is not generally true. It was recently shown~\cite{huang2021information, aharonov2021quantum, huang2021quantum, chen2021hierarchy} that there is an exponential quantum advantage in sample complexity when we can use a quantum sensor, quantum memory, and quantum computer to retrieve, store, and process quantum information from experiments. Such a sample complexity advantage can be proven rigorously without the possibility of being \emph{dequantized}~\cite{tang2019quantum, chia2020sampling, cotler2021revisiting} in the future, i.e., it is impossible to find improved classical algorithms such that there is no exponential advantage.
This significant quantum advantage has recently been demonstrated on the Sycamore processor \cite{huang2021quantum} raising the hope for achieving quantum advantage using NISQ devices \cite{preskill2018quantum}.

The situation for advantage in \emph{time complexity} is more subtle.
Classical simulation of quantum process is intractable in many cases, hence one would expect exponential advantage in time complexity to be prevalent.
However, one should be cautious about the availability of data in ML tasks, which makes classical ML algorithms computationally more powerful \cite{huang2021power, huang2021provably}.
For instance, Ref.~\cite{huang2021provably} shows that \emph{in the worst case}, there is no exponential quantum advantage in predicting ground state properties in geometrically local gapped Hamiltonians.
Furthermore, the emergence of effective classical theory in quantum-mechanical processes could enable classical machines to provide accurate predictions.
For example, density functional theory~\cite{hohenberg1964inhomogeneous, kohn1999nobel} allows accurate prediction of molecular properties when we have an accurate approximation to the exchange-correlation functionals by conducting real-world experiments. 
It is still likely that an exponential advantage is possible in physical systems of practical interest, but there are no rigorous proofs yet.

When the data is of a purely classical origin, such as in applications for recommending products to customers \cite{tang2019quantum}, performing portfolio optimization \cite{alcazar2020classical, bouland2020prospects}, and processing human languages \cite{manning1999foundations} and everyday images \cite{russ2006image}, there is no known exponential advantage \cite{chia2020sampling}.
However, it is still reasonable to expect polynomial advantage. Furthermore, a quadratic advantage can be rigorously proven \cite{grover1996fast, bernstein1997quantum} for purely classical problems.
So we likely have a potential impact in the long-term when we have fault-tolerant quantum computers, albeit with the speedup significantly dampened by the overheads of quantum error correction \cite{babbush2021focus} for currently known fault-tolerant quantum computing schemes.

\subsection{Transition to the fault-tolerant era and beyond}

While QML has been proposed as a candidate to achieve a quantum advantage in the near-term using NISQ devices, one can still pose the question about its usability in the future. Here, researchers envision two different chronological eras post-NISQ. In the first, which we can refer to as ``partial error corrected'', quantum computers will have enough physical qubits (a couple of hundred of them), and sufficiently small error rates, to allow for a small number of fully error corrected logical qubits. Since one logical qubit is comprised of multiple physical qubits, in this era one will have the freedom to trade off and split the qubits in the device onto a subset of error corrected qubits, along with a subset of non-error corrected qubits. The next era, i.e., the ``fault-tolerant era'' will arise when the quantum hardware has a large number of error corrected qubits.

Indeed, one can easily envision QML being useful in both of these post-NISQ  eras. First, in the partial error corrected era, QML models will be able to execute  high-fidelity circuits  and thus have an improved performance. This will naturally enhance the trainability of the models by mitigating noise-induced barren plateaus, and also reduce noise-induced classification errors in QML models. Most importantly, QML will likely see its most widespread and critical use during the fault-tolerant era. Here, quantum algorithms such as those for quantum simulation~\cite{georgescu2014quantum,berry2015simulating} will be able to accurately prepare quantum data, and to faithfully store it in quantum memories~\cite{lvovsky2009optical}. Therefore QML will be the natural model to learn, infer, and make predictions from quantum data, as here the quantum computer will learn from the data itself directly.

On the further-term horizon, we anticipate it will be possible to capture quantum data from nature directly via transduction from its natural analog form to one that is quantum digital (e.g., via quantum analog-digital interconversion \cite{verdon2020qadi}). This data will then be able to be shuttled around quantum networks for distributed and/or centralized processing with quantum machine learning models, using fault-tolerant quantum computation and error-corrected quantum communication. At this point, quantum machine learning will have reached a stage similar to where machine learning is today, where edge sensors capture data, the data is relayed to a central cloud, and machine learning models are trained on the aggregated data. As the modern advent of widespread classical machine learning arose at this point of abundant data, one could anticipate that ubiquitous access to quantum data in the fault-tolerant era could similarly propel quantum machine learning to even greater widespread use.

\section*{Acknowledgements}

MC acknowledges support from the Los Alamos National Laboratory (LANL) LDRD program under project number 20210116DR. MC was also supported by the Center for Nonlinear Studies at LANL. LC and PJC were supported by the U.S. Department of Energy (DOE), Office of Science, Office of Advanced Scientific Computing Research, under the Accelerated Research in Quantum Computing (ARQC) program. LC also acknowledges support from U.S. Department of Energy, Office of Science, National Quantum Information Science Research Centers, Quantum Science Center. PJC was also supported by the NNSA’s Advanced Simulation and Computing Beyond Moore’s Law Program at LANL. GV would like to thank Faris Sbahi, Antonio J. Martinez, and Petar Velickovic for useful discussions. X, formerly known as Google[x], is
part of the Alphabet family of companies, which includes
Google, Verily, Waymo, and others (\url{www.x.company}).
HH is supported by a Google PhD Fellowship.

\section*{Author Contributions}
PJC drafted the manuscript structure. The manuscript was written and revised by MC, GV, HYH, LC and PJC.



\section*{Competing Interests}
The authors declare no competing interests.

\newpage

\end{document}